\documentclass[ALICE,manyauthors]{cernphprep}

\usepackage[comma,square,numbers,sort&compress]{natbib}
\usepackage{hyperref}
\usepackage{lineno}
\begin{document}%

%%%%%%%%%%%%%%%  Title page %%%%%%%%%%%%%%%%%%%%%%%%
\begin{titlepage}
\PHyear{2015}
\PHnumber{069}      % required, will be obtained from PH
\PHdate{13 Mar}  % required, will be obtained from PH
%

%%% Put your own title + short title here:
\title{Search for weakly decaying  $\overline{ \Lambda \mathrm{n}}$ and $ \Lambda \Lambda $ exotic bound states\\in central Pb-Pb collisions at $\mathbf{\sqrt{\textit{s}_{\rm NN}}}$ = 2.76 TeV}
\ShortTitle{Search for weakly decaying dibaryon states}   % appears on right page headers

%%% Do not change the next lines
\Collaboration{ALICE Collaboration\thanks{See Appendix~\ref{app:collab} for the list of collaboration members}}
\ShortAuthor{ALICE Collaboration} % appears on left page headers, do not change

\begin{abstract}
We present results of a search for two hypothetical strange dibaryon states, i.e. the H-dibaryon and the possible $\overline{\Lambda \mathrm{n}}$ bound state. The search is performed with the ALICE detector  in central (0--10\%) Pb--Pb collisions at  $ \sqrt{s_{\mathrm{NN}}} = 2.76 $ TeV,  by invariant mass analysis in the decay modes $\overline{ \Lambda  \mathrm{n}} \rightarrow \overline{\mathrm{d}} \pi^{+} $ and H-dibaryon~$\rightarrow \Lambda \mathrm{p} \pi^{-}$. No evidence for these bound states is observed. Upper limits are determined at 99\% confidence level for a wide range of lifetimes and for the full range of branching ratios. The results are compared to thermal, coalescence and hybrid UrQMD model expectations, which describe correctly the production of other loosely bound states, like the deuteron and the hypertriton. 
\end{abstract}
\end{titlepage}
\setcounter{page}{2}

\section{Introduction}
\label{sec:Intro}
Particle production in Pb--Pb collisions at the Large Hadron Collider (LHC) has been extensively studied \cite{pKpi,pKpi_centrality,lambda_k0s}. The observed production pattern is rather well described in equilibrium thermal models~\cite{thermal_overview,becattini,anton_thermal,cleymans}. Within this approach, the chemical freeze-out temperature $T_{\mathrm{chem}}$, the volume $V$ and the baryo-chemical potential $ \mu _{B}$ are the only three free parameters. 
Even loosely bound states such as the deuteron and hypertriton and their anti-particles have been observed~\cite{star,hypertriton,nuclei} and their rapidity densities are properly described~\cite{pbm1,pbm,thermalModel,thermalModel1,qm2012_pbm,anton1,steinheimer}. 
Consequently other loosely bound states\footnote{The expected masses of these states are some MeV below the sum of the mass of their constituents.} such as the H-dibaryon and the $\Lambda$n are expected to be produced with corresponding yields.

The discovery of the H-dibaryon or the $\Lambda$n bound state would be a breakthrough in hadron spectroscopy as it would imply the existence of a six-quark state and provide crucial information on the $\Lambda$-nucleon and $\Lambda$-$\Lambda$ interaction.
We consequently have started the investigation on the possible existence of such exotic bound states in pp and Pb--Pb collisions at the LHC. Searches for $\Lambda$-nucleon bound states in the $\Lambda$p and $\Lambda$n channels have been carried out (see references~\cite{hires,hires1,hyphi}). 
The H-dibaryon, which is a hypothetical bound state of $uuddss$ $\left(\Lambda\Lambda\right)$, was first predicted by Jaffe using a bag model approach~\cite{jaffe}. Experimental searches have been undertaken since then, but no evidence for a signal was found (see~\cite{h_bnl,belle} and references therein). Recently, the STAR collaboration investigated the $\Lambda$-$\Lambda$ interaction through the measurement of $\Lambda\Lambda$ correlations~\cite{star_h}; this and a theoretical analysis of these data~\cite{kenji} did not reveal a signal.  
Many theoretical investigations of the possible stability of the H-dibaryon have been carried out, but predicting binding energies in the order of MeV for masses of around 2\,GeV/$c^2$ is extremely difficult and challenging~\cite{beane,inoue,shanahan,haidebauer}.

Our approach is to search for such bound states in central Pb--Pb collisions at LHC energies where rapidity densities can be well predicted by thermal~\cite{anton1,steinheimer,rafelski} and coalescence~\cite{exhic} models. %~\cite{steinheimer}. 
The model predictions for rapidity densities of these particles are used and tested against the experimental results.

In this paper the analysis strategies for the searches of the $\overline{ \Lambda \mathrm{n}}\rightarrow \overline{\mathrm{d}} \pi^{+} $ bound state and the H-dibaryon~$\rightarrow \Lambda \mathrm{p} \pi^{-}$ are presented. The analysis focuses on the $\overline{ \Lambda \mathrm{n}}$ bound state because production of anti-particles in the detector material is strongly suppressed and thus secondary contamination of the signal is reduced. 
For the H-dibaryon both the $\Lambda$ and the p originate from secondary vertices where knock-out background is less likely. No search for the anti-H is performed yet, although it is assumed to be produced with equal yield but the measurement depends strongly on the absorption correction.  We begin with a short introduction to the ALICE detector and a description of the particle identification technique used to identify the decay daughters and reconstruct invariant mass distributions. To assess the possible existence of these states we compare the experimental distributions with the model predictions.

\section{Detector setup and data sample}
\label{sec:detector}
The ALICE detector~\cite{alice} is  specifically designed to study heavy-ion collisions. The central barrel comprising the two main tracking detectors, the Inner Tracking System (ITS)~\cite{alice_its} and the Time Projection Chamber (TPC)~\cite{alice_tpc} is housed in a large solenoidal magnet providing a 0.5 T field. The detector pseudorapidity coverage is $\vert \eta \vert \leq$ 0.9 over the full azimuth. An additional part of the central barrel are detectors in forward direction used mainly for triggering and centrality selection. 
The VZERO detectors, two scintillation hodoscopes, are placed on either side of the interaction point and cover the pseudorapidity regions of $2.8<\eta<5.1$ and $-3.7<\eta<-1.7$. The centrality selection is based on the sum of the amplitudes measured in both detectors as described in \cite{vzero2} and \cite{alice_centrality}. \\
The ITS consists of six cylindrical layers of three different types of silicon detectors. The innermost part comprises two silicon pixel (SPD) and two silicon drift detector (SDD) layers. The two outer layers are double-sided silicon microstrip detectors (SSD). Due to the precise space points provided by the ITS a high precision determination of the collision vertex is possible. Therefore, primary and secondary particles can be well separated, down to 100 $\mu$m precision at low transverse momentum ($p_{\mathrm{T}} \approx$ 100\,MeV/$c$).\\
The TPC is the main tracking detector of ALICE and surrounds the ITS. It has a cylindrical design with a diameter of $ \approx $ 550 cm, an inner radius of 85 cm, an outer radius of 247 cm and an overall length in the beam direction of $ \approx $ 510 cm. The 88 m$ ^{3} $ gas volume of the TPC is filled with a mixture of 85.7\% Ne, 9.5\% CO$ _{2} $ and 4.8\% N$ _{2} $. When a charged particle is travelling through the TPC, it ionizes the gas along its path and electrons are released. Due to the uniform electric field along the z-axis (parallel to the beam axis and to the magnetic field) the electrons drift towards the end plates, where the electric signals are amplified and detected in 557568 pads. %With this up to 159 space points can be recorded. 
These data are used to calculate a particle trajectory in the magnetic field and thus determine the track rigidity $ \frac{p}{z} $ (the momentum $p$ of the particle divided by its charge number $z$). The TPC is also used for particle identification via the energy deposit d$E$/d$x$ measurement (see section~\ref{sec:PID}). \\
A complete description of the performance of the ALICE sub-detectors in pp, p--Pb and Pb--Pb collisions can be found in~\cite{alice_performance}.

The searches carried out and reported here are performed by analysing the data set of Pb--Pb collisions from 2011. In the described analyses we use $19.3 \times 10^{6} $ events with a centrality of 0--10\%, determined by the aforementioned VZERO detectors from the previously mentioned campaign.

\section{Particle identification}
\label{sec:PID}
\begin{figure}[th!]
	\centering
	\includegraphics[width=0.8\textwidth]{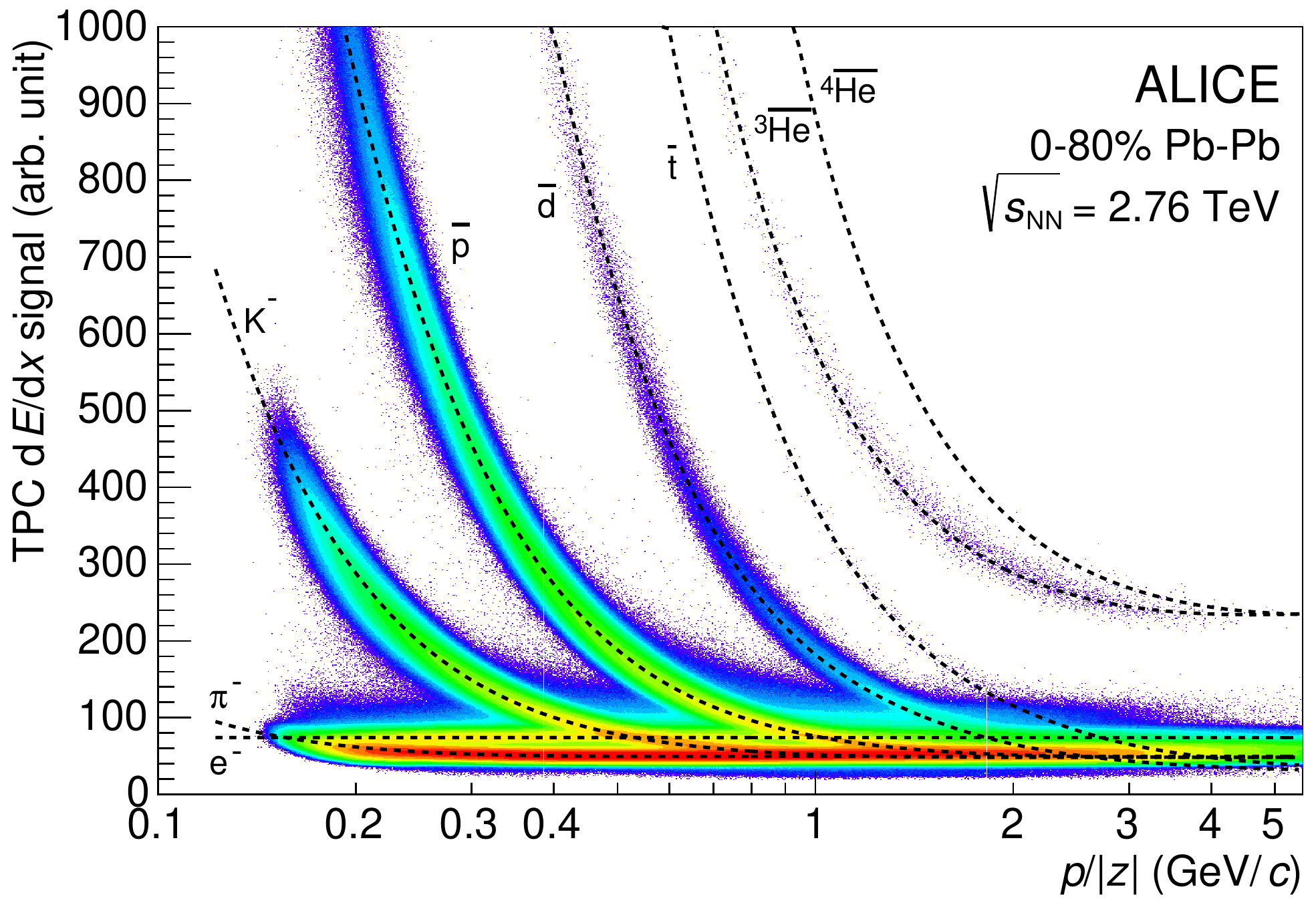} 
	\caption{TPC $ \mathrm{d}E /\mathrm{d}x $ spectrum for negative particles in a sample of three different trigger types (minimum bias, semi-central and central). The dashed lines are parametrisations of the Bethe-Bloch-formula~\cite{bethe,bloch,aleph} for the different particle species.}
	\label{fig:dEdx}
\end{figure}

The precise Particle IDentification (PID) and continuous tracking from very low $ p_{\mathrm{T}}$ (100 MeV/\textit{c}) to moderately high $ p_{\mathrm{T}} $ (20 GeV/\textit{c}) is a unique feature of the ALICE detector at the LHC. The PID used in the analysis described in this letter takes advantage of two different techniques. The energy deposit $\left( \mathrm{d}E/\mathrm{d}x \right) $ and rigidity are measured with the TPC for each reconstructed charged-particle trajectory. This allows the identification of all charged stable particles, from the lightest (electron) to the heaviest ones (anti-alpha). The energy deposit resolution of the TPC  in central Pb--Pb collisions (investigated here) is around 7\%. The corresponding particle separation power is demonstrated in Fig.~\ref{fig:dEdx}. This technique was used in the following to identify the deuterons, protons and pions. The second method makes use of specific topologies from weak decays, which result in typical $V^{0}$ decay patterns. This is used here for the detection of the $\overline{\Lambda\mathrm{n}}$ bound state and the two $V^{0}$ decay patterns of the $\Lambda\Lambda$, namely for the $\Lambda$ identification and the proton--pion decay vertex.    

\section{Analysis}
The strategies of investigation for the two exotic bound states discussed here are quite similar. They both require the detection of a secondary vertex, which in one case is a pure $V^{0}$ and in the second a double $V^{0}$ decay pattern.
We discuss them separately in the following sub-sections. First we describe briefly the common aspects of both analyses.\\
The tracks used in the analyses have to fulfil a set of selection criteria to ensure high tracking efficiency and d$E$/d$x$ resolution. Each track was required to have at least 70 of up to 159 clusters in the TPC attached to it, with the (rather loose) requirement, that the $\chi^2$ of the momentum fit is smaller than 5 per cluster. Tracks with kinks due to weak decays of kaons and pions are rejected. To achieve final precision the accepted tracks are refit while the track finding algorithm is run inwards, outwards and inwards again (for more details on the ALICE tracking see \cite{alice_performance} and section 5 of~\cite{PPR}).\\
$V^{0}$ decays are determined by two (or more) tracks which are emitted from a secondary vertex and which might come close to each other (the minimum distance is called Distance-of-Closest-Approach DCA) while each of the tracks has a certain minimum distance (DCA of the track to a vertex) to the primary vertex. A powerful selection criterion for detecting proper $V^{0}$ candidates is the restriction of the pointing angle, namely the angle between the reconstructed flight-line and the reconstructed momentum of the $V^{0}$ particle. More details of the secondary vertex reconstruction can be found in~\cite{alice_performance,PPR,lambda_k0s}, where also the clear and effective identification of $\Lambda$ baryons is displayed using the aforementioned technique.  
The selection criteria, described below, are optimised using a Monte Carlo set where the simulated exotic bound states are assumed to live as long as a free $\Lambda$ baryon. This is a reasonable assumption for all strange dibaryons, which are expected to live around 2--4$\times 10^{-10}$ s~\cite{dilambda,PhysRevLett.71.1328,Jurgen2} in the regions of binding energies investigated here.  

\begin{table}[th!]
  \centering
  \begin{tabular}{l | c }
    {\bf Selection criterion} & {\bf Value} \\
    \hline
    Track selection criteria &  \\
    \hline
    Tracks with kinks & rejected \\
    Number of clusters in TPC & $n_{\mathrm{cl}} > 70$ \\
    Track quality & $\chi^{2}/\mathrm{cluster} < 5$ \\
    Acceptance in pseudorapidity   & $ \vert \eta \vert < 0.9 $ \\
    Acceptance in rapidity & $ \vert y \vert < 1 $\\
    \hline
    $V^{0}$ and kinematic selection criteria &  \\
    \hline
    Pointing angle &  $\Theta < 0.045$\,rad          \\ 
    DCA between the $V^{0}$ daughters &  DCA $< 0.3 \; \mathrm{cm}$    \\
    Momentum $ p_{\mathrm{tot}}$ of the anti-deuteron &  $ p_{\mathrm{tot}} > 0.2 \; \mathrm{GeV/}c $\\
    Energy deposit d$E$/d$x$ anti-deuteron&  d$E$/d$x > 110 $ (from Fig.~\ref{fig:dEdx})\\ 
    PID cut for daughters &  $\pm$3$\sigma $ (TPC)\\ 
    \hline
  \end{tabular}
  \caption{Selection criteria for $\overline{\Lambda\mathrm{n}}$ analysis.}
  \label{tab:cuts anti-lambdaN}
\end{table}

\begin{figure}[th!]
	\centering
       \includegraphics[width=0.8\textwidth]{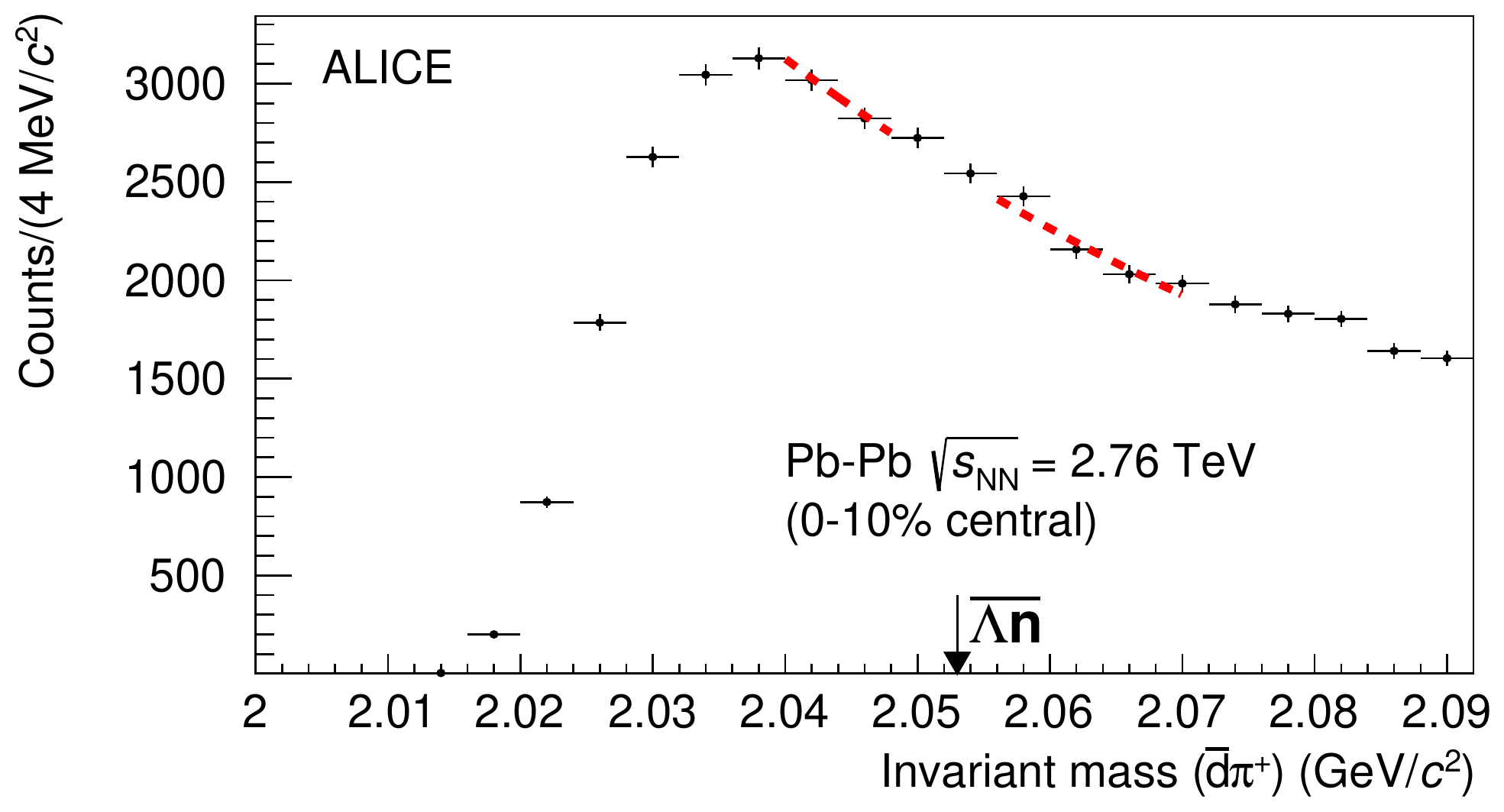}
	\caption{Invariant mass distribution for $\overline{\mathrm{d}}\pi^{+}$ for the Pb--Pb data corresponding to $ 19.3 \times 10^{6}$ central events. The arrow indicates the sum of the mass of the constituents ($\overline{\Lambda\mathrm{n}}$) of the assumed bound state. A signal for the bound state is expected in the region below this sum. The dashed line represents an exponential fit outside the expected signal region to estimate the background.}
	\label{fig:InvaMass_AntiLambdaN}
\end{figure} 

\subsection{$\overline{ \Lambda \mathrm{n}}$ bound state}
In analogy to recent hypertriton measurements~\cite{hypertriton,star} we focus here on the expected two-body decay $\overline{ \Lambda  \mathrm{n}} \rightarrow \overline{\mathrm{d}} \pi^{+}$. 
For the data analysis the following strategy is used: first displaced vertices are identified using ITS and TPC information. In a second step the negative track of the $V^{0}$ candidate is identified as an anti-deuteron via the TPC $\mathrm{d}E/\mathrm{d}x $ information. If the second daughter is identified as a pion, the invariant mass of the pair is reconstructed. Both particles are required to lie within a 3 standard  deviations ($\sigma$) band of the expected Bethe-Bloch lines of the corresponding particles.  
To identify the secondary vertex the two daughter tracks have to have a DCA smaller than 0.3\,cm. Another condition is that the maximum pointing angle is smaller than 0.045 rad (see description above). Deuterons are cleanly identified in the rigidity region of 400 MeV/$c$ to 1.75 GeV/$c$. To limit contamination from other particle species, the d$E$/d$x$ has to be above 110 units of the TPC signal, shown in Fig.~1.\\
The selection criteria are summarised in Table~\ref{tab:cuts anti-lambdaN}.  
The resulting invariant mass distribution, reflecting the kinematic range of identified daughter tracks, is displayed in Fig.~\ref{fig:InvaMass_AntiLambdaN}.

\subsection{ H-dibaryon}
The search for the H-dibaryon is performed in the decay channel $ \mathrm{H} \rightarrow \Lambda \mathrm{p}\pi^{-} $, with a  mass lying in the range $ 2.200 \, \mathrm{GeV}/c^{2}<m_{\mathrm{H}}<2.231 \, \mathrm{GeV}/c^{2} $ (see Fig.~\ref{fig:InvaMass_Hdibaryon} below). The analysis strategy for the H-dibaryon is similar as for the $ \overline{\Lambda \mathrm{n}}$ bound state described above, except that here a second $V^{0}$-type decay particle is involved.\\
One $V^{0}$ candidate originating from the H-dibaryon decay vertex has to be identified as a $\Lambda$ decaying into a proton and a pion. In addition another $V^{0}$ decay pattern reconstructed from a proton and a pion is required to be found at the decay vertex of the H-dibaryon. First the invariant mass of the $\Lambda$ is reconstructed and then the candidates in the
invariant mass window of $ 1.111\,\mathrm{GeV}/c^2 < m_{\Lambda} < 1.120\,\mathrm{GeV}/c^2$ are combined with the four-vectors of the proton and pion at the decay vertex. %The decay vertex is constructed via the two opposite signed tracks. 
A 3$\sigma$ d$E$/d$x$ cut in the TPC is used to identify the protons and the pions for both the $\Lambda$ candidate and the $V^{0}$ topology at the H-dibaryon decay vertex.\\ 
 To cope with the huge background caused by primary and secondary pions additional selection criteria have to be applied.  Each track is required to be at least 2\,cm away from the primary vertex and the tracks combined to a $V^{0}$ are required to have a minimum distance below 1\,cm. The pointing angle is required to be below 0.05\,rad. %0.9987
All selection criteria are summarised in Table~\ref{tab_new_cuts_h}. 
The resulting invariant mass is shown in Fig.~\ref{fig:InvaMass_Hdibaryon}. The shape of the invariant mass distribution is caused by the kinematic range of the identified daughter tracks.

\begin{table}[h!]
  \centering
  \begin{tabular}{l | c }
    {\bf Selection criterion} & {\bf Value} \\
    \hline
    Track selection criteria &  \\
    \hline
    Tracks with kinks  & rejected \\
    Number of clusters in TPC & $n_{\mathrm{cl}} > 80$ \\
    Track quality & $\chi^{2}/\mathrm{cluster} < 5$ \\
    Acceptance in pseudorapidity  & $ \vert \eta \vert < 0.9 $ \\
    Acceptance in rapidity & $ \vert y \vert < 1 $\\
    \hline
    $V^{0}$ selection criteria &  \\
    \hline
    DCA $V^{0}$ daughters & DCA $< 1$\,cm \\
    DCA positive $V^{0}$ daughter - H decay vertex & DCA $> 2$\,cm \\
    DCA negative $V^{0}$ daughter - H decay vertex & DCA $> 2$\,cm \\
    \hline
    Kinematic selection criteria &  \\    
    \hline
    DCA positive H daughter -  Primary vertex & DCA $> 2$\,cm \\
    DCA negative H daughter - Primary vertex & DCA $> 2$\,cm \\
    DCA H daughters & DCA $< 1$\,cm \\
    Pointing angle of H & $\Theta < 0.05$\,rad \\
    PID cut for daughters &  $\pm$3$\sigma $ (TPC)\\ 
    $\Lambda$ mass window  &  $\pm$3$\sigma$\\ 
    \hline
  \end{tabular}
  \caption{Selection criteria used for $\Lambda\Lambda$ (H-dibaryon) analysis.}
  \label{tab_new_cuts_h}
\end{table}

\begin{figure}[th!]
	\centering
\includegraphics[width=0.8\textwidth]{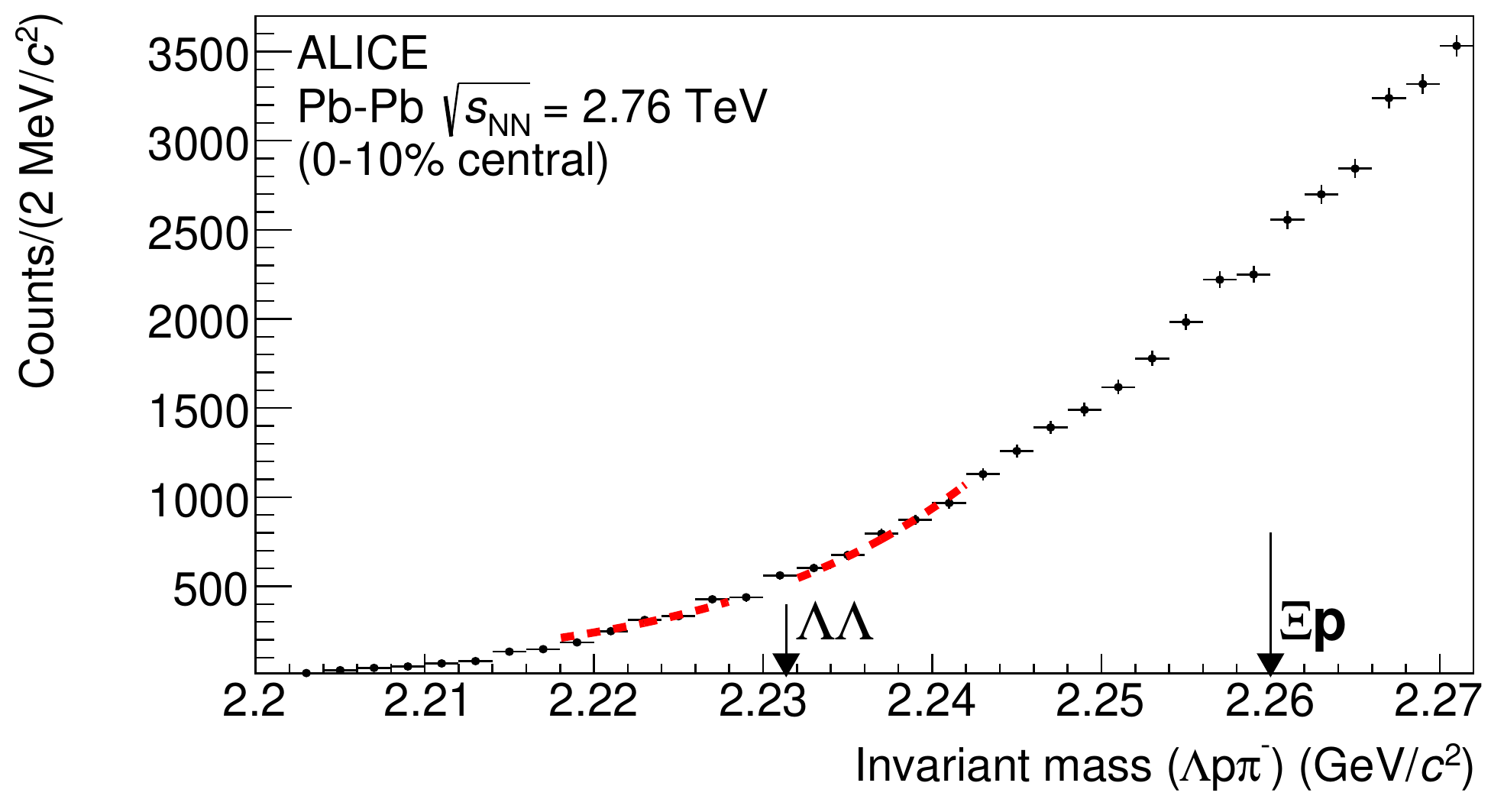}
	\caption{Invariant mass distribution for $\Lambda \mathrm{p} \pi^{-} $ for the Pb--Pb data corresponding to $ 19.3 \times 10^{6}$ central events. The left arrow indicates the sum of the masses of the constituents ($\Lambda\Lambda$) of the possible bound state. A signal for the bound state is expected in the region below this sum. For the speculated resonant state a signal is expected between the $\Lambda\Lambda$ and the $\Xi$p (indicated by the right arrow) thresholds. The dashed line is an exponential fit to estimate the background.}
	\label{fig:InvaMass_Hdibaryon}
\end{figure}

\section{Systematics and absorption correction}
\label{systematics}

Monte Carlo samples have been produced to estimate the efficiency for the detection of the $ \overline{\Lambda \mathrm{n}}$ bound state and the H-dibaryon. The kinematical distributions of the hypothetical bound states were generated uniformly in rapidity $y$ and in transverse momentum $p_{\mathrm{T}}$. In order to deal with the unknown lifetime, different decay lengths are investigated, ranging from 4 cm up to 3 m. The lower limit is determined by the secondary vertex finding efficiency and the upper limit by the requirement that there is a significant probability for decays inside the TPC\footnote{For the H-dibaryon there is also a theoretical maximal decay length calculated for the investigated decay channel~\cite{donoghue}.} (the final acceptance$\times$efficiency drops down to 1\% for the $ \overline{\Lambda \mathrm{n}}$ and $10^{-3}$ for the H-dibaryon).  
The shape of transverse momentum spectra in heavy-ion collisions is described well by the blast-wave approach, with radial flow parameter $ \langle \beta \rangle$ and kinetic freeze-out temperature $T_\mathrm{kin}$ as in~\cite{centrality_production}. The true shape of the $ p_{\mathrm{T}}$ spectrum is also not known, therefore it is estimated from the extrapolation of blast-wave fits to deuterons and $ ^{3}\mathrm{He}$ spectra at the same energy~\cite{nuclei}. To obtain final efficiencies, the resulting blast-wave distributions constructed for the exotic bound states are normalised to unity and convoluted with the correction factors (efficiency $ \times $ acceptance). 

Typical values of the final efficiency are of the order of a few percent assuming the lifetime of the free $\Lambda$. The uncertainty in the shape of the $ p_{\mathrm{T}}$ distributions is the main source of systematic error. Blast-wave fits of deuteron and $^{3}\mathrm{He}$ spectra are employed to explore the range of systematic uncertainties. Analyses of these results lead to a systematic uncertainty in the overall yield of around 25\%.

Other systematic uncertainties are estimated by varying the cuts described in Table~\ref{tab:cuts anti-lambdaN} and Table~\ref{tab_new_cuts_h} within the limits consistent with the detector resolution. The contributions of these systematic uncertainties are typically found to be in the percent range. The combination of the different sources leads to a global systematic uncertainty of around 30\% for both analyses, when all uncertainties are added in quadrature.

For the $\overline{\Lambda \mathrm{n}}$ bound state analysis the possible absorption of the anti-deuterons and the bound state itself when crossing material has to be taken into account. For this, the same procedure as used for the anti-hypertriton analysis~\cite{hypertriton} is utilised.
The absorption correction ranges from 3 to 40\% (depending on the lifetime of the $ \overline{\Lambda \mathrm{n}} $ bound state, which determines the amount of material crossed) with an overall uncertainty of 7\%. 

\section{Results}
No significant signal in the invariant mass distributions has been observed for both cases, as visible from~Fig.~\ref{fig:InvaMass_AntiLambdaN} and Fig.~\ref{fig:InvaMass_Hdibaryon}\footnote{Note that a hypothetical H-dibaryon with a mass above the $\Xi$p threshold would not be observable in the present analysis.}.   
The shape of the invariant mass distribution of $\overline{\mathrm{d}}\pi^{+}$ is of purely kinematic origin, reflecting the momentum distribution of the particles used.
The selection criteria listed in Table~\ref{tab:cuts anti-lambdaN} are tuned to select secondary decays. 
The secondary anti-deuterons involved in the analysis originate mainly from two sources: The first and dominating source are daughters from three-body decays of the anti-hypertriton ($^{3}_{\bar{\Lambda}}\overline{\mathrm{H}}\rightarrow\bar{\mathrm{d}}\bar{\mathrm{p}}\pi^{+}$ and $^{3}_{\bar{\Lambda}}\overline{\mathrm{H}}\rightarrow\bar{\mathrm{d}}\bar{\mathrm{n}}\pi^{0}$) where the other decay daughters are not detected. The invariant mass spectrum is obtained by combining theses anti-deuterons with pions generated in the collision. The second source is due to prompt anti-deuterons which are incorrectly labelled as displaced, because they have such low momenta that the DCA resolution of these tracks is not sufficient to separate primary from secondary particles. 

Since no signal in the invariant mass distributions is observed upper limits are estimated. For the estimation of upper limits for the rapidity density d$N$/d$y$ the method discussed in~\cite{rolke} is utilised. In particular, we apply the software package $TRolke$ as implemented in $ROOT$~\cite{root}.  This method needs as input mass and experimental width (3$\sigma$) of the hypothetical bound states. The observed counts are therefore compared to a smooth background as given by an exponential fit outside the signal region (as indicated by the line in Fig.~\ref{fig:InvaMass_AntiLambdaN} and Fig.~\ref{fig:InvaMass_Hdibaryon}). For both candidates  $\overline{\Lambda \mathrm{n}}$ and H-dibaryon we assume a binding energy of 1 MeV. The width is determined by the experimental resolution and obtained from Monte Carlo simulations. In addition, the final efficiency which is discussed in section~\ref{systematics} is required.
Further, values of branching ratios of the assumed bound states are needed. These depend strongly on the binding energy. With a 1 MeV binding energy for the $\overline{\Lambda \mathrm{n}}$ bound state the branching ratio in the $\overline{\mathrm{d}}+\pi^{+}$ decay channel is expected to be 54\%~\cite{Jurgen1}. 
The branching ratio for a 1 MeV or less bound H-dibaryon decaying into $\Lambda\mathrm{p}\pi^{-}$ is predicted to be 64\%, see~\cite{Jurgen2}. 

\begin{figure}[th!]
	\centering
	\includegraphics[width=0.8\textwidth]{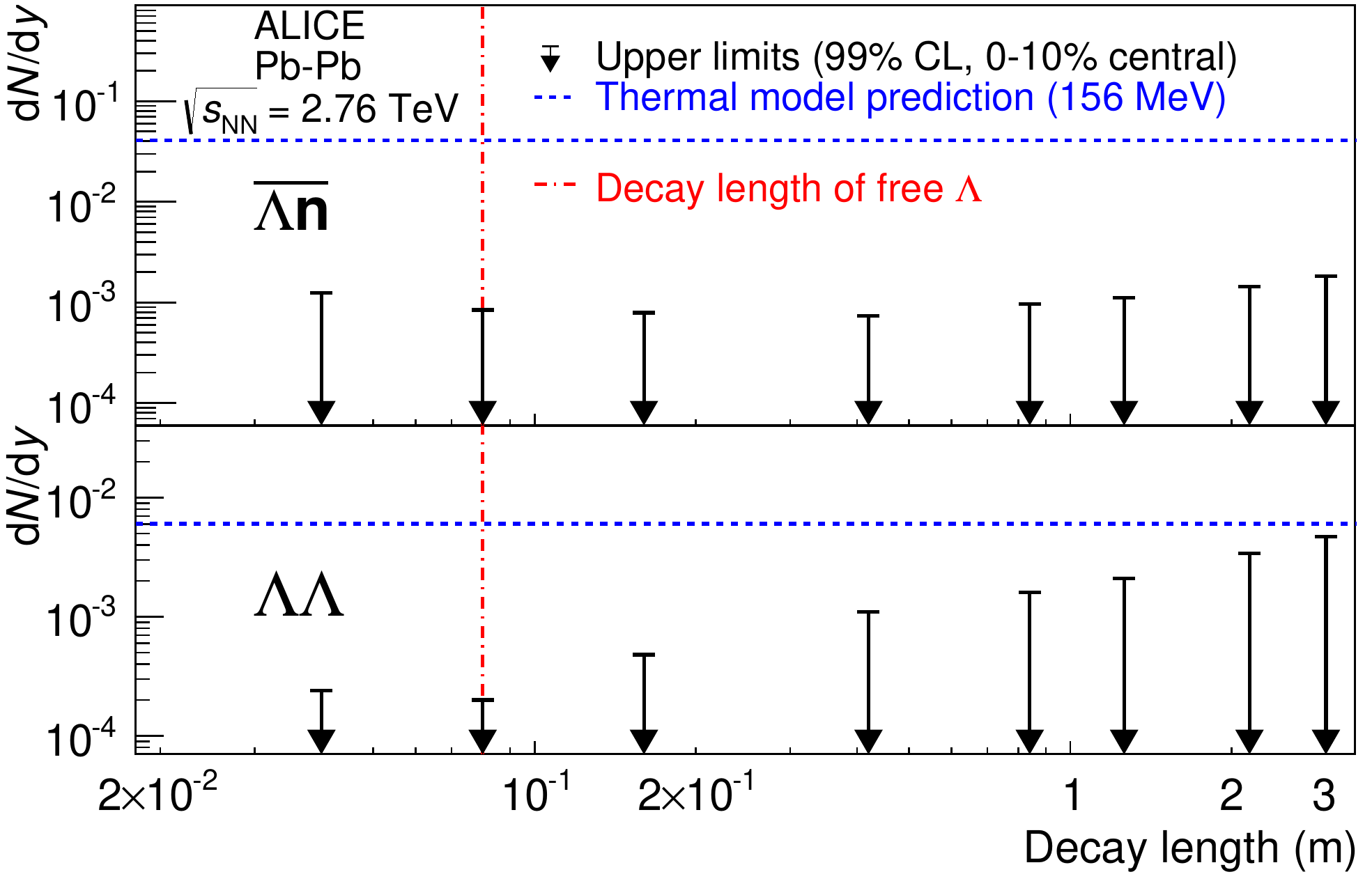}
	\caption{Upper limit of the rapidity density as function of the decay length shown for the $\overline{\Lambda \mathrm{n}}$ bound state in the upper panel and for the H-dibaryon in the lower panel. Here a branching ratio of 64\% was used for the H-dibaryon and a branching ratio of 54\% for the $\overline{\Lambda \mathrm{n}}$ bound state. The horizontal (dashed) lines indicate the expectation of the thermal model with a temperature of 156 MeV. The vertical line shows the lifetime of the free $\Lambda$ baryon.}
		\label{fig:UpperLimitsLifetime}
\end{figure}

The resulting upper limits, for 99\% CL, are shown in Fig.~\ref{fig:UpperLimitsLifetime} as a function of the different lifetimes; for the $\overline{\Lambda \mathrm{n}}$ bound state in the upper panel and for the H-dibaryon in the lower panel. These upper limits include systematic uncertainties. For the $\overline{\Lambda \mathrm{n}}$ the absorption corrections are also considered in the figure, which causes the upper limits to be shifted upwards. 

The obtained upper limits can now be compared to model predictions. The rapidity densities $ \mathrm{d}N/ \mathrm{d}y $ from a thermal model prediction for a chemical freeze-out temperature of, for example, 156 MeV, are $ \mathrm{d}N/ \mathrm{d}y = 4.06 \times 10^{-2}$ for the $ \overline{\Lambda \mathrm{n}} $ bound state and $ \mathrm{d}N/ \mathrm{d}y = 6.03 \times 10^{-3}$ for the H-dibaryon~\cite{anton1}. These values are indicated with the (blue) dashed lines in Fig.~\ref{fig:UpperLimitsLifetime}. 
For the investigated range of lifetimes the upper limit of the $\overline{\Lambda \mathrm{n}}$ bound state is at least a factor 20 below this prediction.
For the H-dibaryon the upper limits depend more strongly on the lifetime since it has a different decay topology and all four final state tracks have to be reconstructed. The upper limit is a factor of 20 below the thermal model prediction for the lifetime of the free $\Lambda$ and becomes less stringent at higher lifetimes since the detection efficiency becomes small. For a lifetime of 10$^{-8}$\,s, corresponding to a decay length of 3\,m, the difference between model and upper limit reduces to a factor two. 

\begin{figure}[th!]
	\centering
    \includegraphics[width=0.8\textwidth]{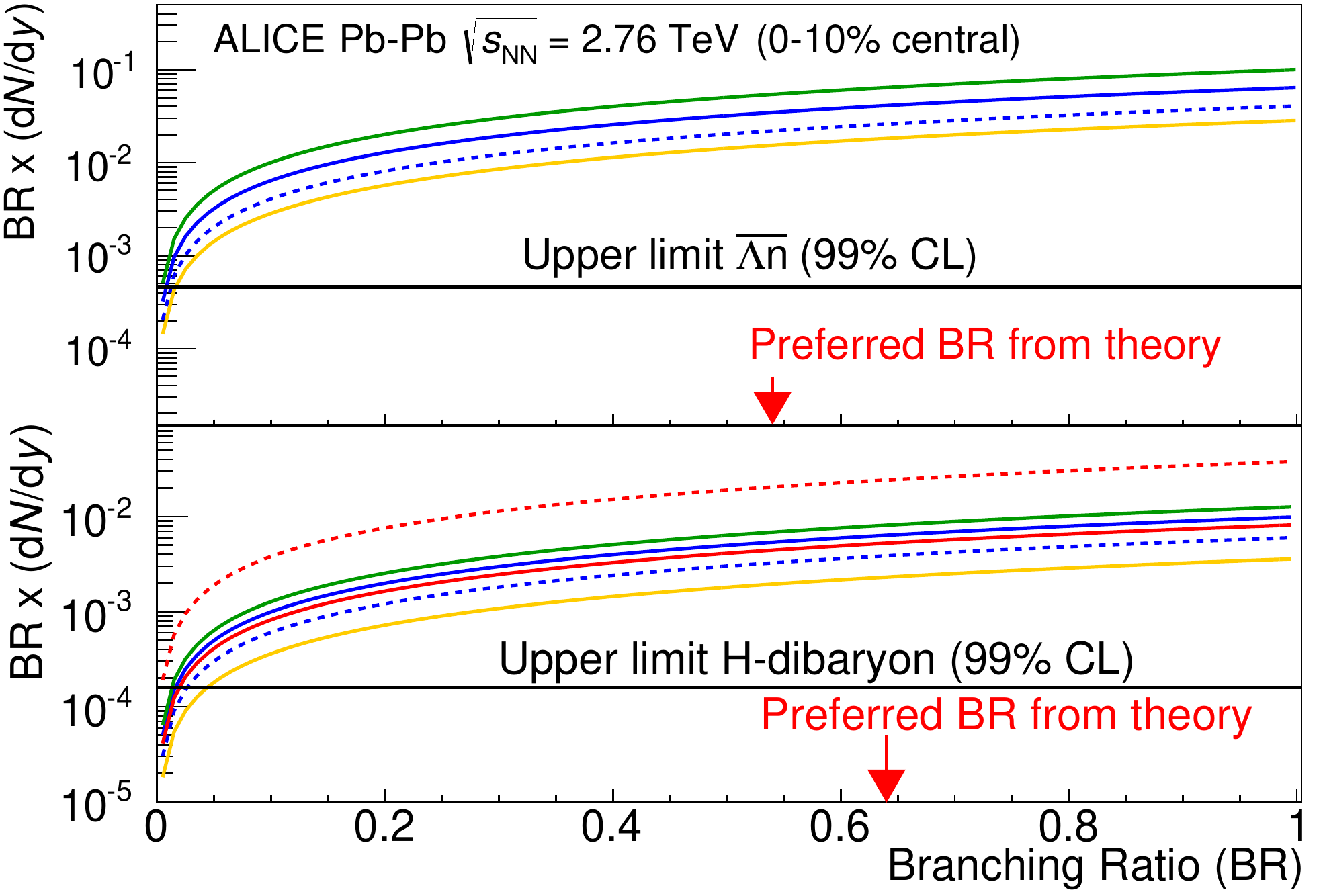} 
	\caption{Experimentally determined upper limit, under the assumption of the lifetime of a free $\Lambda$. In the upper panel shown for the $\overline{\Lambda \mathrm{n}}$ bound state and for the H-dibaryon in the lower panel. It includes 30\% systematic uncertainty for each particle and 6\% correction for absorption with an uncertainty of 7\% for the $\overline{\Lambda \mathrm{n}}$ bound state. The theory lines are drawn for different theoretical branching ratios (BR) in blue for the equilibrium thermal model from~\cite{anton1} for two temperatures (164 MeV the full line and 156 MeV the dashed line), in green the non-equilibrium thermal model from~\cite{rafelski} and in yellow the predictions from a hybrid UrQMD calculation~\cite{jan}. The H-dibaryon is also compared with predictions from coalescence models, where the full red line visualises the prediction assuming quark coalescence and the dashed red line corresponds to hadron coalescence~\cite{exhic}.}
	\label{fig:UpperLimitBR_AntiLambdaN}
\end{figure}

In order to take the uncertainties in the branching ratio into account, we plot in Fig.~\ref{fig:UpperLimitBR_AntiLambdaN} the products of the upper limit of the rapidity density times the branching ratio together with several theory predictions~\cite{anton1,rafelski,exhic,jan}. 
The curves are obtained using the value for the $\Lambda$-lifetime of Fig.~\ref{fig:UpperLimitsLifetime}. 

The (red) arrows in the figures indicate the branching ratio from the theory predictions~\cite{Jurgen1,Jurgen2}.  The obtained upper limits are a factor of more than 5 below all theory predictions for a branching ratio of at least 5\% for the $\overline{\Lambda \mathrm{n}}$ bound state and at least 20\% for the H-dibaryon. 

\section{Discussion}
The limits obtained on the rapidity density of the investigated exotic compound objects are found to be more than one order of magnitude below the expectations of particle production models, when using a realistic branching ratio and a reasonable lifetime.
It has to be noted that simultaneously, a clear signal was observed for the very loosely bound hypertriton (binding energy $<150$\,keV) for which production yields have been measured~\cite{hypertriton}. These yields along with those of nuclei A~=~2,3,4 agree well with the predictions of the thermal model discussed above and decrease with each additional baryon number by roughly a factor 300. 
One would therefore assume that the yield of the $\overline{\Lambda\mathrm{n}}$, if such particle existed, should also be predicted by this model and with a value for the rapidity density of about a factor 300 higher than the measured hypertriton yield. Similar considerations hold for the H-dibaryon.

\section{Conclusion}
A search is reported for the existence of loosely bound strange dibaryons $\Lambda\Lambda$ and $\overline{ \Lambda \mathrm{n}}$ whose possible existence has been discussed widely in the literature. No signals are observed. On the other hand, loosely bound objects with baryon number A~=3 such as the hypertriton have been measured in the same data sample. 
The yields of nuclei~\cite{nuclei} and of the hypertriton~\cite{hypertriton} are quantitatively understood within a thermal model calculation.
The present analysis provides stringent upper limits at 99\% confidence level for the production of H-dibaryon and $\Lambda\mathrm{n}$ bound state, in general significantly below the thermal model predictions. The upper limits are obtained for different lifetimes. The values are well below the model predictions when realistic branching ratios and reasonable lifetimes are assumed. Thus, our results do not support the existence of the H-dibaryon and the $\Lambda\mathrm{n}$ bound state. 

               %%%%%%%%%%% put the body of the article here
%
%

%%%%% acknowledgements
\newenvironment{acknowledgement}{\relax}{\relax}
\begin{acknowledgement}
\section*{Acknowledgements}
We thank S.~Beane, M. Petr\'{a}\v{n}, J.~Schaffner-Bielich and J.~Steinheimer for useful correspondence.\\
% $Id: acknowledgements.tex 2098 2015-04-24 15:54:16Z loizides $
% Version: Jan 2015

The ALICE Collaboration would like to thank all its engineers and technicians for their invaluable contributions to the construction of the experiment and the CERN accelerator teams for the outstanding performance of the LHC complex.
The ALICE Collaboration gratefully acknowledges the resources and support provided by all Grid centres and the Worldwide LHC Computing Grid (WLCG) collaboration.
The ALICE Collaboration acknowledges the following funding agencies for their support in building and
running the ALICE detector:
State Committee of Science,  World Federation of Scientists (WFS)
and Swiss Fonds Kidagan, Armenia,
Conselho Nacional de Desenvolvimento Cient\'{\i}fico e Tecnol\'{o}gico (CNPq), Financiadora de Estudos e Projetos (FINEP),
Funda\c{c}\~{a}o de Amparo \`{a} Pesquisa do Estado de S\~{a}o Paulo (FAPESP);
National Natural Science Foundation of China (NSFC), the Chinese Ministry of Education (CMOE)
and the Ministry of Science and Technology of China (MSTC);
Ministry of Education and Youth of the Czech Republic;
Danish Natural Science Research Council, the Carlsberg Foundation and the Danish National Research Foundation;
The European Research Council under the European Community's Seventh Framework Programme;
Helsinki Institute of Physics and the Academy of Finland;
French CNRS-IN2P3, the `Region Pays de Loire', `Region Alsace', `Region Auvergne' and CEA, France;
German Bundesministerium fur Bildung, Wissenschaft, Forschung und Technologie (BMBF) and the Helmholtz Association;
General Secretariat for Research and Technology, Ministry of
Development, Greece;
Hungarian Orszagos Tudomanyos Kutatasi Alappgrammok (OTKA) and National Office for Research and Technology (NKTH);
Department of Atomic Energy and Department of Science and Technology of the Government of India;
Istituto Nazionale di Fisica Nucleare (INFN) and Centro Fermi -
Museo Storico della Fisica e Centro Studi e Ricerche "Enrico
Fermi", Italy;
MEXT Grant-in-Aid for Specially Promoted Research, Ja\-pan;
Joint Institute for Nuclear Research, Dubna;
National Research Foundation of Korea (NRF);
Consejo Nacional de Cienca y Tecnologia (CONACYT), Direccion General de Asuntos del Personal Academico(DGAPA), M\'{e}xico, Amerique Latine Formation academique - European Commission~(ALFA-EC) and the EPLANET Program~(European Particle Physics Latin American Network);
Stichting voor Fundamenteel Onderzoek der Materie (FOM) and the Nederlandse Organisatie voor Wetenschappelijk Onderzoek (NWO), Netherlands;
Research Council of Norway (NFR);
National Science Centre, Poland;
Ministry of National Education/Institute for Atomic Physics and National Council of Scientific Research in Higher Education~(CNCSI-UEFISCDI), Romania;
Ministry of Education and Science of Russian Federation, Russian
Academy of Sciences, Russian Federal Agency of Atomic Energy,
Russian Federal Agency for Science and Innovations and The Russian
Foundation for Basic Research;
Ministry of Education of Slovakia;
Department of Science and Technology, South Africa;
Centro de Investigaciones Energeticas, Medioambientales y Tecnologicas (CIEMAT), E-Infrastructure shared between Europe and Latin America (EELA), Ministerio de Econom\'{i}a y Competitividad (MINECO) of Spain, Xunta de Galicia (Conseller\'{\i}a de Educaci\'{o}n),
Centro de Aplicaciones Tecnológicas y Desarrollo Nuclear (CEA\-DEN), Cubaenerg\'{\i}a, Cuba, and IAEA (International Atomic Energy Agency);
Swedish Research Council (VR) and Knut $\&$ Alice Wallenberg
Foundation (KAW);
Ukraine Ministry of Education and Science;
United Kingdom Science and Technology Facilities Council (STFC);
The United States Department of Energy, the United States National
Science Foundation, the State of Texas, and the State of Ohio;
Ministry of Science, Education and Sports of Croatia and  Unity through Knowledge Fund, Croatia.
Council of Scientific and Industrial Research (CSIR), New Delhi, India
    %%%%%%% done by webmaster team
\end{acknowledgement}

%%%%%%%% Bibliography (In case of using bibtex generate the bbl requested by arXiv)
\bibliographystyle{utphys}   % Put here the style file name for the paper, i.e.apsrev4-1, utphys
%\bibliography{biblio}
\bibliography{bib_paper_exotica}
%\input {bibliography.tex}  

%%%%%%%%% appendix with author list
\newpage
\appendix
\section{The ALICE Collaboration}
\label{app:collab}

% Collaboration: CERN-LHC-ALICE
% Generation Date is 2015/Jan/14

% How to use:
%%%%%%%%% appendix with author list
%\appendix
%\section{The ALICE Collaboration}
%\label{app:collab}
%\input{authors-list.tex}  %%%%%%% get the latest version before submitting

\begingroup
\small
\begin{flushleft}
J.~Adam\Irefn{org39}\And
D.~Adamov\'{a}\Irefn{org82}\And
M.M.~Aggarwal\Irefn{org86}\And
G.~Aglieri Rinella\Irefn{org36}\And
M.~Agnello\Irefn{org110}\And
N.~Agrawal\Irefn{org47}\And
Z.~Ahammed\Irefn{org130}\And
I.~Ahmed\Irefn{org16}\And
S.U.~Ahn\Irefn{org67}\And
I.~Aimo\Irefn{org93}\textsuperscript{,}\Irefn{org110}
% ...
\renewcommand\labelenumi{\textsuperscript{\theenumi}~}

\section*{Affiliation notes}
\renewcommand\theenumi{\roman{enumi}}
\begin{Authlist}
\item \Adef{0}Deceased
\item \Adef{idp26511132}{Also at: University of Kansas, Lawrence, Kansas, United States}
\end{Authlist}

\section*{Collaboration Institutes}
\renewcommand\theenumi{\arabic{enumi}~}
\begin{Authlist}

\item \Idef{org1}A.I. Alikhanyan National Science Laboratory (Yerevan Physics Institute) Foundation, Yerevan, Armenia
\item \Idef{org2}Benem\'{e}rita Universidad Aut\'{o}noma de Puebla, Puebla, Mexico
\item \Idef{org3}Bogolyubov Institute for Theoretical Physics, Kiev, Ukraine
\item \Idef{org4}Bose Institute, Department of Physics and Centre for Astroparticle Physics and Space Science (CAPSS), Kolkata, India
\item \Idef{org5}Budker Institute for Nuclear Physics, Novosibirsk, Russia
\item \Idef{org6}California Polytechnic State University, San Luis Obispo, California, United States
\item \Idef{org7}Central China Normal University, Wuhan, China
\item \Idef{org8}Centre de Calcul de l'IN2P3, Villeurbanne, France
\item \Idef{org9}Centro de Aplicaciones Tecnol\'{o}gicas y Desarrollo Nuclear (CEADEN), Havana, Cuba
\item \Idef{org10}Centro de Investigaciones Energ\'{e}ticas Medioambientales y Tecnol\'{o}gicas (CIEMAT), Madrid, Spain
\item \Idef{org11}Centro de Investigaci\'{o}n y de Estudios Avanzados (CINVESTAV), Mexico City and M\'{e}rida, Mexico
\item \Idef{org12}Centro Fermi - Museo Storico della Fisica e Centro Studi e Ricerche ``Enrico Fermi'', Rome, Italy
\item \Idef{org13}Chicago State University, Chicago, Illinois, USA
\item \Idef{org14}China Institute of Atomic Energy, Beijing, China
\item \Idef{org15}Commissariat \`{a} l'Energie Atomique, IRFU, Saclay, France
\item \Idef{org16}COMSATS Institute of Information Technology (CIIT), Islamabad, Pakistan
\item \Idef{org17}Departamento de F\'{\i}sica de Part\'{\i}culas and IGFAE, Universidad de Santiago de Compostela, Santiago de Compostela, Spain
\item \Idef{org18}Department of Physics and Technology, University of Bergen, Bergen, Norway
\item \Idef{org19}Department of Physics, Aligarh Muslim University, Aligarh, India
\item \Idef{org20}Department of Physics, Ohio State University, Columbus, Ohio, United States
\item \Idef{org21}Department of Physics, Sejong University, Seoul, South Korea
\item \Idef{org22}Department of Physics, University of Oslo, Oslo, Norway
\item \Idef{org23}Dipartimento di Elettrotecnica ed Elettronica del Politecnico, Bari, Italy
\item \Idef{org24}Dipartimento di Fisica dell'Universit\`{a} 'La Sapienza' and Sezione INFN Rome, Italy
\item \Idef{org25}Dipartimento di Fisica dell'Universit\`{a} and Sezione INFN, Cagliari, Italy
\item \Idef{org26}Dipartimento di Fisica dell'Universit\`{a} and Sezione INFN, Trieste, Italy
\item \Idef{org27}Dipartimento di Fisica dell'Universit\`{a} and Sezione INFN, Turin, Italy
\item \Idef{org28}Dipartimento di Fisica e Astronomia dell'Universit\`{a} and Sezione INFN, Bologna, Italy
\item \Idef{org29}Dipartimento di Fisica e Astronomia dell'Universit\`{a} and Sezione INFN, Catania, Italy
\item \Idef{org30}Dipartimento di Fisica e Astronomia dell'Universit\`{a} and Sezione INFN, Padova, Italy
\item \Idef{org31}Dipartimento di Fisica `E.R.~Caianiello' dell'Universit\`{a} and Gruppo Collegato INFN, Salerno, Italy
\item \Idef{org32}Dipartimento di Scienze e Innovazione Tecnologica dell'Universit\`{a} del  Piemonte Orientale and Gruppo Collegato INFN, Alessandria, Italy
\item \Idef{org33}Dipartimento Interateneo di Fisica `M.~Merlin' and Sezione INFN, Bari, Italy
\item \Idef{org34}Division of Experimental High Energy Physics, University of Lund, Lund, Sweden
\item \Idef{org35}Eberhard Karls Universit\"{a}t T\"{u}bingen, T\"{u}bingen, Germany
\item \Idef{org36}European Organization for Nuclear Research (CERN), Geneva, Switzerland
\item \Idef{org37}Faculty of Engineering, Bergen University College, Bergen, Norway
\item \Idef{org38}Faculty of Mathematics, Physics and Informatics, Comenius University, Bratislava, Slovakia
\item \Idef{org39}Faculty of Nuclear Sciences and Physical Engineering, Czech Technical University in Prague, Prague, Czech Republic
\item \Idef{org40}Faculty of Science, P.J.~\v{S}af\'{a}rik University, Ko\v{s}ice, Slovakia
\item \Idef{org41}Faculty of Technology, Buskerud and Vestfold University College, Vestfold, Norway
\item \Idef{org42}Frankfurt Institute for Advanced Studies, Johann Wolfgang Goethe-Universit\"{a}t Frankfurt, Frankfurt, Germany
\item \Idef{org43}Gangneung-Wonju National University, Gangneung, South Korea
\item \Idef{org44}Gauhati University, Department of Physics, Guwahati, India
\item \Idef{org45}Helsinki Institute of Physics (HIP), Helsinki, Finland
\item \Idef{org46}Hiroshima University, Hiroshima, Japan
\item \Idef{org47}Indian Institute of Technology Bombay (IIT), Mumbai, India
\item \Idef{org48}Indian Institute of Technology Indore, Indore (IITI), India
\item \Idef{org49}Inha University, Incheon, South Korea
\item \Idef{org50}Institut de Physique Nucl\'eaire d'Orsay (IPNO), Universit\'e Paris-Sud, CNRS-IN2P3, Orsay, France
\item \Idef{org51}Institut f\"{u}r Informatik, Johann Wolfgang Goethe-Universit\"{a}t Frankfurt, Frankfurt, Germany
\item \Idef{org52}Institut f\"{u}r Kernphysik, Johann Wolfgang Goethe-Universit\"{a}t Frankfurt, Frankfurt, Germany
\item \Idef{org53}Institut f\"{u}r Kernphysik, Westf\"{a}lische Wilhelms-Universit\"{a}t M\"{u}nster, M\"{u}nster, Germany
\item \Idef{org54}Institut Pluridisciplinaire Hubert Curien (IPHC), Universit\'{e} de Strasbourg, CNRS-IN2P3, Strasbourg, France
\item \Idef{org55}Institute for Nuclear Research, Academy of Sciences, Moscow, Russia
\item \Idef{org56}Institute for Subatomic Physics of Utrecht University, Utrecht, Netherlands
\item \Idef{org57}Institute for Theoretical and Experimental Physics, Moscow, Russia
\item \Idef{org58}Institute of Experimental Physics, Slovak Academy of Sciences, Ko\v{s}ice, Slovakia
\item \Idef{org59}Institute of Physics, Academy of Sciences of the Czech Republic, Prague, Czech Republic
\item \Idef{org60}Institute of Physics, Bhubaneswar, India
\item \Idef{org61}Institute of Space Science (ISS), Bucharest, Romania
\item \Idef{org62}Instituto de Ciencias Nucleares, Universidad Nacional Aut\'{o}noma de M\'{e}xico, Mexico City, Mexico
\item \Idef{org63}Instituto de F\'{\i}sica, Universidad Nacional Aut\'{o}noma de M\'{e}xico, Mexico City, Mexico
\item \Idef{org64}iThemba LABS, National Research Foundation, Somerset West, South Africa
\item \Idef{org65}Joint Institute for Nuclear Research (JINR), Dubna, Russia
\item \Idef{org66}Konkuk University, Seoul, South Korea
\item \Idef{org67}Korea Institute of Science and Technology Information, Daejeon, South Korea
\item \Idef{org68}KTO Karatay University, Konya, Turkey
\item \Idef{org69}Laboratoire de Physique Corpusculaire (LPC), Clermont Universit\'{e}, Universit\'{e} Blaise Pascal, CNRS--IN2P3, Clermont-Ferrand, France
\item \Idef{org70}Laboratoire de Physique Subatomique et de Cosmologie, Universit\'{e} Grenoble-Alpes, CNRS-IN2P3, Grenoble, France
\item \Idef{org71}Laboratori Nazionali di Frascati, INFN, Frascati, Italy
\item \Idef{org72}Laboratori Nazionali di Legnaro, INFN, Legnaro, Italy
\item \Idef{org73}Lawrence Berkeley National Laboratory, Berkeley, California, United States
\item \Idef{org74}Lawrence Livermore National Laboratory, Livermore, California, United States
\item \Idef{org75}Moscow Engineering Physics Institute, Moscow, Russia
\item \Idef{org76}National Centre for Nuclear Studies, Warsaw, Poland
\item \Idef{org77}National Institute for Physics and Nuclear Engineering, Bucharest, Romania
\item \Idef{org78}National Institute of Science Education and Research, Bhubaneswar, India
\item \Idef{org79}Niels Bohr Institute, University of Copenhagen, Copenhagen, Denmark
\item \Idef{org80}Nikhef, National Institute for Subatomic Physics, Amsterdam, Netherlands
\item \Idef{org81}Nuclear Physics Group, STFC Daresbury Laboratory, Daresbury, United Kingdom
\item \Idef{org82}Nuclear Physics Institute, Academy of Sciences of the Czech Republic, \v{R}e\v{z} u Prahy, Czech Republic
\item \Idef{org83}Oak Ridge National Laboratory, Oak Ridge, Tennessee, United States
\item \Idef{org84}Petersburg Nuclear Physics Institute, Gatchina, Russia
\item \Idef{org85}Physics Department, Creighton University, Omaha, Nebraska, United States
\item \Idef{org86}Physics Department, Panjab University, Chandigarh, India
\item \Idef{org87}Physics Department, University of Athens, Athens, Greece
\item \Idef{org88}Physics Department, University of Cape Town, Cape Town, South Africa
\item \Idef{org89}Physics Department, University of Jammu, Jammu, India
\item \Idef{org90}Physics Department, University of Rajasthan, Jaipur, India
\item \Idef{org91}Physik Department, Technische Universit\"{a}t M\"{u}nchen, Munich, Germany
\item \Idef{org92}Physikalisches Institut, Ruprecht-Karls-Universit\"{a}t Heidelberg, Heidelberg, Germany
\item \Idef{org93}Politecnico di Torino, Turin, Italy
\item \Idef{org94}Purdue University, West Lafayette, Indiana, United States
\item \Idef{org95}Pusan National University, Pusan, South Korea
\item \Idef{org96}Research Division and ExtreMe Matter Institute EMMI, GSI Helmholtzzentrum f\"ur Schwerionenforschung, Darmstadt, Germany
\item \Idef{org97}Rudjer Bo\v{s}kovi\'{c} Institute, Zagreb, Croatia
\item \Idef{org98}Russian Federal Nuclear Center (VNIIEF), Sarov, Russia
\item \Idef{org99}Russian Research Centre Kurchatov Institute, Moscow, Russia
\item \Idef{org100}Saha Institute of Nuclear Physics, Kolkata, India
\item \Idef{org101}School of Physics and Astronomy, University of Birmingham, Birmingham, United Kingdom
\item \Idef{org102}Secci\'{o}n F\'{\i}sica, Departamento de Ciencias, Pontificia Universidad Cat\'{o}lica del Per\'{u}, Lima, Peru
\item \Idef{org103}Sezione INFN, Bari, Italy
\item \Idef{org104}Sezione INFN, Bologna, Italy
\item \Idef{org105}Sezione INFN, Cagliari, Italy
\item \Idef{org106}Sezione INFN, Catania, Italy
\item \Idef{org107}Sezione INFN, Padova, Italy
\item \Idef{org108}Sezione INFN, Rome, Italy
\item \Idef{org109}Sezione INFN, Trieste, Italy
\item \Idef{org110}Sezione INFN, Turin, Italy
\item \Idef{org111}SSC IHEP of NRC Kurchatov institute, Protvino, Russia
\item \Idef{org112}SUBATECH, Ecole des Mines de Nantes, Universit\'{e} de Nantes, CNRS-IN2P3, Nantes, France
\item \Idef{org113}Suranaree University of Technology, Nakhon Ratchasima, Thailand
\item \Idef{org114}Technical University of Split FESB, Split, Croatia
\item \Idef{org115}The Henryk Niewodniczanski Institute of Nuclear Physics, Polish Academy of Sciences, Cracow, Poland
\item \Idef{org116}The University of Texas at Austin, Physics Department, Austin, Texas, USA
\item \Idef{org117}Universidad Aut\'{o}noma de Sinaloa, Culiac\'{a}n, Mexico
\item \Idef{org118}Universidade de S\~{a}o Paulo (USP), S\~{a}o Paulo, Brazil
\item \Idef{org119}Universidade Estadual de Campinas (UNICAMP), Campinas, Brazil
\item \Idef{org120}University of Houston, Houston, Texas, United States
\item \Idef{org121}University of Jyv\"{a}skyl\"{a}, Jyv\"{a}skyl\"{a}, Finland
\item \Idef{org122}University of Liverpool, Liverpool, United Kingdom
\item \Idef{org123}University of Tennessee, Knoxville, Tennessee, United States
\item \Idef{org124}University of the Witwatersrand, Johannesburg, South Africa
\item \Idef{org125}University of Tokyo, Tokyo, Japan
\item \Idef{org126}University of Tsukuba, Tsukuba, Japan
\item \Idef{org127}University of Zagreb, Zagreb, Croatia
\item \Idef{org128}Universit\'{e} de Lyon, Universit\'{e} Lyon 1, CNRS/IN2P3, IPN-Lyon, Villeurbanne, France
\item \Idef{org129}V.~Fock Institute for Physics, St. Petersburg State University, St. Petersburg, Russia
\item \Idef{org130}Variable Energy Cyclotron Centre, Kolkata, India
\item \Idef{org131}Vin\v{c}a Institute of Nuclear Sciences, Belgrade, Serbia
\item \Idef{org132}Warsaw University of Technology, Warsaw, Poland
\item \Idef{org133}Wayne State University, Detroit, Michigan, United States
\item \Idef{org134}Wigner Research Centre for Physics, Hungarian Academy of Sciences, Budapest, Hungary
\item \Idef{org135}Yale University, New Haven, Connecticut, United States
\item \Idef{org136}Yonsei University, Seoul, South Korea
\item \Idef{org137}Zentrum f\"{u}r Technologietransfer und Telekommunikation (ZTT), Fachhochschule Worms, Worms, Germany
\end{Authlist}
\endgroup

  %%%%%%% done by webmaster team
\end{document}